\documentclass[preprint,superscriptaddress,showkeys]{revtex4}

\usepackage{amsmath}
\usepackage{amssymb}
\usepackage[dvips]{graphicx}

\newcommand{\ket}[1]{\left\vert#1\right\rangle}
\newcommand{\bra}[1]{\left\langle#1\right\vert}

\begin{document}
\title{Density Matrix Renormalization Group for Dummies}

\author{Gabriele De Chiara}
\affiliation{CNR-INFM-BEC \& Dipartimento di Fisica,
  Universita di Trento, I-38050 Povo, Italy}

\author{Matteo Rizzi}
\email[Corresponding author: ]{m.rizzi@sns.it}
\affiliation{NEST-CNR-INFM \& Scuola Normale Superiore,
  Piazza dei Cavalieri 7, I-56126 Pisa, Italy}

\author{Davide Rossini}
\affiliation{NEST-CNR-INFM \& Scuola Normale Superiore,
  Piazza dei Cavalieri 7, I-56126 Pisa, Italy}

\author{Simone Montangero}
\affiliation{NEST-CNR-INFM \& Scuola Normale Superiore,
  Piazza dei Cavalieri 7, I-56126 Pisa, Italy}


\date{\today}
\begin{abstract}
We describe the Density Matrix Renormalization Group algorithms for 
time dependent and time independent Hamiltonians. This paper is a 
brief but comprehensive introduction to the subject for anyone willing 
to enter in the field or write the program source code from scratch. 
An open source version of the code can be found at:
\texttt{http://qti.sns.it/dmrg/phome.html} .
\end{abstract}

\maketitle


The advent of information era has been opening the possibility to 
perform numerical simulations of quantum many-body systems, thus 
revealing completely new perspectives in the field of condensed matter 
theory. Indeed, together with the analytic approaches, numerical 
techniques provide a lot of information and details otherwise 
inaccessible. However, the simulation of a quantum mechanical system 
is generally a very hard task; one of the main reasons is related to 
the number of parameters required to represent a quantum state. This 
value usually grows exponentially with the number of constituents of 
the system,$^{1}$ due to the corresponding exponential 
growth of the Hilbert space. This exponential scaling drastically 
reduces the possibility of a direct simulation of many-body quantum 
systems. In order to overcome this limitation, many numerical tools 
have been developed, such as Monte Carlo techniques$^{2}$ or 
efficient Hamiltonian diagonalization methods, like Lanczos and Davidson 
procedures.$^{3}$

The Density Matrix Renormalization Group (DMRG) method has been 
introduced by White in 1992.$^{4,5}$ It was originally 
devised as a numerical algorithm useful for simulating ground state 
properties of one-dimensional quantum lattices, such as the Heisenberg 
model or Bose-Hubbard models; then it has also been adapted in order 
to simulate small two-dimensional systems.$^{6,7}$
DMRG traces its roots to 
Wilson's numerical Renormalization Group (RG),$^{8}$ which 
represents the simplest way to perform a real-space renormalization of 
Hamiltonians. Starting from a numerical representation of some 
microscopic Hamiltonian in a particular basis, degrees of freedom are 
iteratively added, typically by increasing the size of the finite 
system. Then less important ones are integrated out and accounted for 
by modifying the original Hamiltonian. The new Hamiltonian will thus 
exhibit modified as well as new couplings; renormalization group 
approximations consist in physically motivated truncations of the set 
of couplings newly generated by the elimination of degrees of freedom. 
In this way one obtains a simplified effective Hamiltonian that should 
catch the essential physics of the system under study.
We point out that the DMRG can also be seen as a variational method
under the matrix-product-form ansatz for trial wave functions:
the ground state and elementary excited states in the thermodynamic limit
can be simply expressed via an ansatz form which can be explored
variationally, without referencing to the renormalization
construction.$^{9}$
Very recently, influence from the quantum information community
has led to a DMRG-like algorithm which is able to simulate the temporal
evolution of one-dimensional quantum systems.$^{10,11,12,13,14,15,16,17}$

Quantum information theory has also allowed to clarify the situations 
in which this method can be applied efficiently. Indeed, it has been 
shown$^{10}$ that the efficiency in simulating a quantum 
many-body system is strictly connected to its entanglement behavior. 
More precisely, if the entanglement of a subsystem with respect to the 
whole is bounded (or grows logarithmically  with its size) an 
efficient simulation with DMRG is possible. Up to now, it is known that ground 
states of one dimensional lattices (whether critical or not) satisfy 
this requirement, whereas in higher dimensionality it is not fulfilled 
as the entanglement is subject to an area law.$^{18}$
On the 
other hand, the simulation of the time evolution of critical systems 
may not be efficient even in one dimensional systems as the block 
entanglement can grow linearly with time and block size.$^{19,20}$
In a different context, it has also 
been shown that in a quantum computer performing an efficient quantum 
algorithm (Shor's algorithm and the simulation of a quantum chaotic 
system) the entanglement between qubits grows faster than 
logarithmically.$^{21,22}$
Thus, t-DMRG cannot efficiently simulate every quantum one dimensional
system; 
nonetheless, its range of applicability is very broad and embraces 
very different subjects. Indeed, DMRG can be used to study 
condensed matter problems and to simulate many quantum information 
applications in 1D quantum systems as,
for example, simulations of quantum information transfer,$^{23}$ 
quantum computations,$^{21}$ the effects of decoherence on a
qubit$^{24}$ and the entanglement properties of one dimensional
critical systems.$^{25}$ 

The aim of this review is to introduce the reader to the last 
development of DMRG codes, briefly but in a comprehensive way. For the 
sake of briefness we do not review the vast literature of papers based 
on DMRG techniques and we refer the interested readers to
Refs.~${6,7,26}$ and references therein. 
Here we provide both the main ideas and the technicalities needed to 
reach a deep understanding of DMRG and allowing an interested reader 
to develop its own DMRG code or modify an existing one. We also 
provide some easy examples that can be used as testbeds for new DMRG 
codes. 

In Sec.~\ref{sec:staticDMRG} we describe the basics of time independent 
DMRG algorithm, in Sec.~\ref{sec:measDMRG} we introduce the 
measurement procedure (a more detailed exposition is given in 
Ref.~$7$ and references therein). In 
Sec.~\ref{sec:timeDMRG} the time dependent DMRG algorithm is 
explained. Finally, in Sec.~\ref{sec:examples} we provide some 
numerical examples, and in Sec.~\ref{sec:techDMRG} we discuss some 
technical issues regarding the implementation of a DMRG program code. 
In the last section the reader can find the schemes of the DMRG 
algorithms, both for the static and time dependent case. Further 
material can be found on our website 
(\texttt{http://qti.sns.it/dmrg/phome.html}), where 
the t-DMRG code will be released with an open license.

\section{The static DMRG algorithm}\label{sec:staticDMRG}

As yet pointed out in the introduction, the tensorial structure of the 
Hilbert space of a composite system leads to an exponential growth of 
the resources needed for the simulation with the number of the system 
constituents. However, if one is interested in the ground state properties
of a one-dimensional system, the number of parameters is limited for
non critical systems or grows polynomially for a critical one.$^{18}$
This implies that it is possible to rewrite the 
state of the system in a more efficient way, i.e. it can be described 
by using a number of coefficients which is much smaller than the 
dimension of the Hilbert space. Equivalently, a strategy to simulate 
ground state properties of a system is to consider only a relevant 
subset of states of the full Hilbert space. This idea is at the heart
of the so called ``real space blocking renormalizarion group''
which we briefly describe below, and is reminiscent of the
renormalization group (RG) introduced by Wilson.$^{8}$

In the real space blocking RG procedure one typically begins
with a small part of a 
quantum system (a block $\mathcal{B}$ of size $L$, living on an 
$m$-dimensional Hilbert space), and a Hamiltonian which describes the 
interaction between two identical blocks. Then one projects the 
composite 2-block system (of size $2L$) representation (dimension 
$m^2$) onto the subspace spanned by the $m$ lowest-lying energy 
eigenstates, thus obtaining a new truncated representation for it. 
Each operator is consequently projected onto the new $m$-dimensional 
basis. This procedure is then iteratively repeated, until the desired 
system size is reached. RG was successfully applied for the Kondo 
problem, but fails in the description of strongly interacting systems. 
This failure is due to the procedure followed to increase the system 
size and to the criterion used to select the representative states of 
the renormalized block: indeed the decimation procedure of the Hilbert 
space is based on the assumption that the ground state of the entire 
system will essentially be composed of energetically low-lying states 
living on smaller subsystems (the forming blocks) which is not always 
true. {A simple counter-example is given by a free particle 
in a box: the ground state with length $2 l$ has no nodes, whereas any 
combination of two grounds in $l$ boxes will have a node in the 
middle, thus resulting in higher energy.}

A convenient strategy to solve the RG breakdown is the following: 
before choosing the states to be retained for a finite-size block, it 
is first embedded in some environment that mimics the thermodynamic 
limit of the system. This is the new key ingredient of the DMRG 
algorithm; the price one has to pay is a slowdown of the system growth 
with the number of the algorithm's iterations: from the exponentially 
fast growth Wilson's procedure to the DMRG linear growth
(very recently, in the context of real-space renormalization group methods,
a new scheme which recovers the exponential growth has been proposed;
this is based upon a coarse-graining transformation that renormalizes
the amount of entanglement of a block before its truncation$^{27}$).
In the following, we introduce the working 
principles of the DMRG, and provide a detailed description to 
implement it in practice (for a pedagogical introduction see for 
example Refs.~$28,29$).

\subsection{Infinite-system DMRG}\label{ssec:infiniteDMRG}

Keeping in mind the main ideas of the DMRG depicted above, we now 
formulate the basis structure of the so called \emph{infinite-system} 
DMRG for one-dimensional lattice systems.%
The typical scenario where DMRG can be used is the search for an 
approximate ground state of a 1D chain of neighbor interacting 
\emph{sites}, each of them living in a Hilbert space of dimension $D$. 
As in Wilson's RG, DMRG is an iterative procedure in which the system 
is progressively enlarged. In the infinite system algorithm we keep 
enlarging the system until the ground state properties we are 
interested in (e.g., the ground state energy per site) have converged. 

The system Hamiltonian is written as: 
\begin{equation} \label{htot}
\hat{H} = \sum_i \sum_q J(q) \hat{S}_i(q)\hat{T}_{i+1}(q) +
\hat{B}(q) \hat{V}_i(q)
\end{equation}
where $J(q)$ and $B(q)$ are coupling constants, and $\{ \hat{S}_i(q) 
\}_q$, $\{ \hat{T}_i(q)\}_q$ and $\{ \hat{V}_i(q) \}_q$ are sets of 
operators acting on the $i$-th site. The index $q$ refers to the 
various elements of these sets. For example, in a magnetic chain these 
can be angular momentum operators. For simplicity we will not describe 
the case of position dependent couplings, since it can be easily 
reduced to the uniform case.

\begin{figure}[!ht]
  \begin{center}
    \includegraphics[scale=0.6]{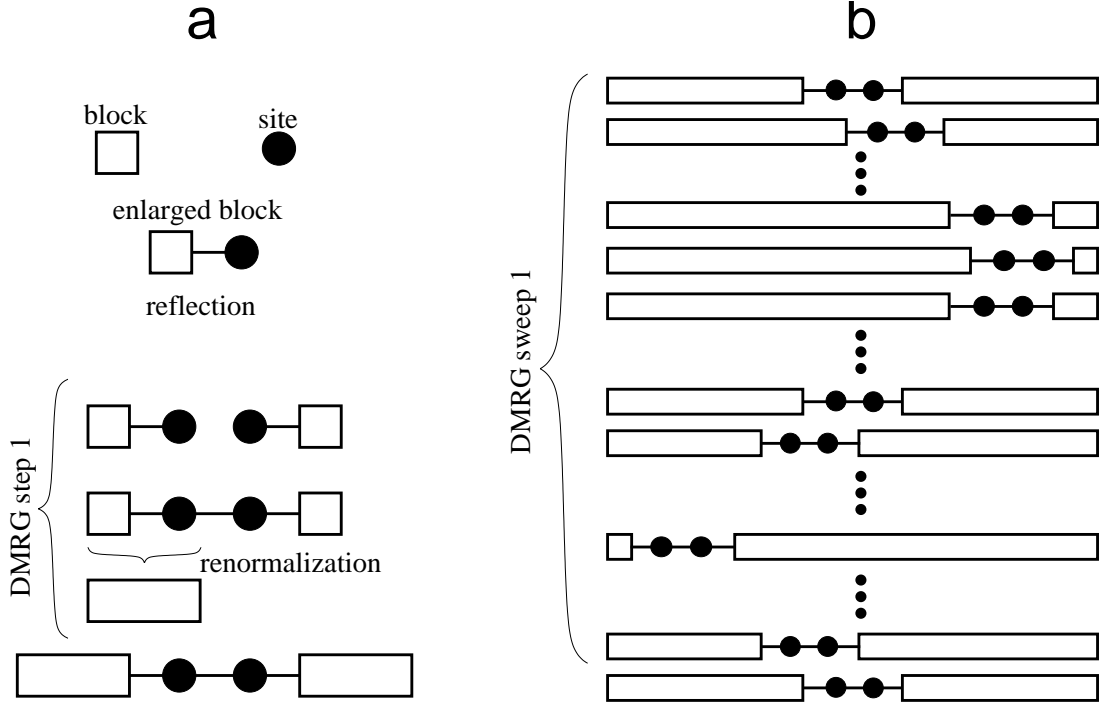}
    \caption{Schematic procedure for the DMRG algorithm.
      On the left part (a) one iteration of the infinite-system DMRG
      algorithm is shown:
      starting from the system block $\mathcal B(L,m_L)$ and adding one free
      site to it, the enlarged block $\mathcal B(L,m_L) \, \bullet$ is formed.
      Here for simplicity we assume that the system is reflection-symmetric,
      thus the environmental right block is taken equal to the left block.
      Then, after having created the super-block
      $\mathcal B(L,m_L) \, \bullet \, \bullet \, \mathcal B(L,m_L)$,
      a renormalization procedure is applied in order to get the new block
      for the next DMRG iteration. \newline
      On the right part (b) the scheme of a complete finite-system DMRG sweep
      is depicted.}
    \label{fig:blocks}
  \end{center}
\end{figure}

The algorithm starts with a \emph{block} composed of one site 
$\mathcal B(1,D)$ (see Fig.~\ref{fig:blocks}a); the arguments of 
$\mathcal B$ refer to the number of sites it embodies, and to the 
number of states used to describe it. From the computational point of 
view, a generic block $\mathcal B(L,m_L)$ is a portion of memory which 
contains all the information about the block: the block Hamiltonian, 
its basis and other operators that we will introduce later. The block 
Hamiltonian $\hat{H}_B$ for $\mathcal B(L,m_L)$ includes only the 
local terms (i.e. local and interaction terms where only sites 
belonging to the block are involved). The next step consists in 
building the so called left \emph{enlarged block}, by adding a site to 
the right of the previously created block. The corresponding 
Hamiltonian $\hat{H}_E$ is composed by the local Hamiltonians of the 
block and the site, plus the interaction term:
\begin{equation}
\hat{H}_E = \hat{H}_B + \hat{H}_S + \hat{H}_{BS} \; .
\end{equation}
The enlarged block is then coupled to a similarly constructed right 
enlarged block. If the system has global reflection symmetry, the 
right enlarged block Hamiltonian $\hat{H}_{E'}$ can be obtained just 
by reflecting the left enlarged block.$^{30}$

By adding the interaction of the two enlarged blocks, a \emph{super-block} 
Hamiltonian $\hat{H}_{supB}$ is then built, which describes the global 
system:
\begin{equation}
\hat{H}_{supB} = \hat{H}_E + \hat{H}_{E'} + \hat{H}_{SS'} \; .
\end{equation}
From now on, we refer to the sites $S$ and $S'$ as the \emph{free 
sites}. The matrix $\hat{H}_{supB}$ should finally be diagonalized in 
order to find the ground state $\psi_G$, which can be rewritten in ket 
notation as: 
\begin{equation}
\ket{\psi_G}=\psi_{a\alpha\beta b}\ket{a\alpha\beta b}.
\end{equation}
Hereafter Latin indexes refer to blocks, while Greek indexes indicate
free sites; implicit summation convention is assumed. From 
$\ket{\psi_G}$ one evaluates the reduced density matrix $\hat{\rho}_L$ 
of the left enlarged block, by tracing out the right enlarged block:
\begin{equation}
\hat{\rho}_L = \textrm{Tr}_R\ket{\psi_G}\bra{\psi_G} = 
\psi_{a\alpha\beta b} \, \psi^*_{a'\alpha'\beta b} \ket{a \, \alpha} 
\bra{a' \alpha'} \, .
\end{equation}
The core of the DMRG algorithm stands in the renormalization procedure 
of the enlarged block, which eventually consists in finding a 
representation in terms of a reduced basis with at most $m$ (fixed 
\emph{a priori}) elements. This corresponds to a truncation of the 
Hilbert space of the enlarged block, since $m_{L+1} = \min (m_L D, m)$.$^{31}$
These states are chosen to be the first 
$m_{L+1}$ eigenstates of $\rho_L$, corresponding to the largest 
eigenvalues. This truncated change of basis is performed by using the 
$m_L D \times m_{L+1}$ rectangular matrix $\hat{O}_{L\to L+1}$ (where 
the subscripts stand for the number of sites enclosed in the input 
block and in the output renormalized block), whose {columns}, 
in matrix representation, are the $m_{L+1}$ selected eigenstates. To 
simplify notations, let us introduce the function 
$g(a,\alpha)=D(a-1)+\alpha$, which acts on a block index $a$ and on 
the next free site index $\alpha$ and gives an index of the enlarged 
block running from $1$ to $m_L D$. The output of the full 
renormalization procedure is a truncated enlarged block $\mathcal 
B(L+1,m_{L+1})$, which coincides with the new starting block for the 
next DMRG iteration. This consists in the new block Hamiltonian:
\begin{eqnarray}	 
\hat{H}_{B}' & = & \hat{O}^{\dagger}_{L\to L+1} \:\Hat{H}_E \: 
\hat{O}_{L\to L+1} =\\ \nonumber
& = & O^{\! *\, g(a,\alpha) \, c}_{L\to L+1} \: 
H^{g(a,\alpha)\, g(a',\alpha')}_E O^{g(a',\alpha') \, c'}_{L\to L+1} 
\, \ket{c}\bra{c'}
\end{eqnarray}
and in the local operators:
\begin{equation}\label{eq:updateS}
\hat{S}'_{L+1}(q) = \hat{O}^\dagger_{L\to L+1} \: \hat{S}_{L+1}(q) \: 
\hat{O}_{L\to L+1}
\end{equation}
written in the new basis. These are necessary in the next step, for 
the construction of the interaction between the rightmost block site 
and the free site. The output block $\mathcal B(L+1,m_{L+1})$ includes 
also the matrix $\hat{O}_{L\to L+1}$ which identifies the basis states 
of the new block.

It is worth to emphasize that we can increase the size of our system 
without increasing the number of states describing it, by iteratively 
operating the previously described procedure.

We now summarize the key operations needed to perform a single DMRG 
step. For each DMRG step the dimension of the super-block Hamiltonian 
goes from $2L$ to $2L +2$, thus the simulated system size increases by 
$2$ sites. The infinite-system DMRG, with reflection symmetry, 
consists in iterating these operations:
\begin{enumerate}
\item{Start from left block $\mathcal B(L,m_L)$, and enlarge it by 
adding the interaction with a single site.} \item{Reflect such 
enlarged block, in order to form the right enlarged block.} 
\item{Build the super-block from the interaction of the two enlarged 
blocks.}
\item{Find the ground state of the super-block and the 
$m_{L+1} = \min (m_L D, m)$ eigenstates of the reduced density matrix 
of the left enlarged block with largest eigenvalues.} 
\item{Renormalize all the relevant operators with the matrix 
$\hat{O}_{L\to L+1}$, thus obtaining $\mathcal B(L+1,m_{L+1})$.}
\end{enumerate} 
Notice that at each DMRG step the ground state of a chain whose length
grows by two sites is found. By contrast, the number of states describing a
block is always $m$, regardless of how many sites it includes.
This means that the complexity of the problem is a priori fixed by $m$ and $D$
(while $D$ is imposed by the structure of the simulated system,
$m \geq D$ is a parameter which has to be appropriately set up by the user,
in order to get the desired precision for the simulation; see also
Sec.~\ref{sec:examples}).
In Sec.~\ref{sec:techDMRG} we will discuss how it is possible to extract the
ground state of the super-block Hamiltonian without finding
its entire spectrum, by means of efficient numerical diagonalization methods,
like Davidson or Lanczos algorithms.$^{32}$
We stress that at each DMRG step a truncation error $\epsilon_{\mathrm{tr}}$
is introduced:
\begin{equation}
\epsilon_{\mathrm{tr}} = \sum_{i>m} \lambda_i
\end{equation}
where $\lambda_i$ are the eigenvalues of the reduced density matrix
$\rho_L$ in decreasing order.
The error $\epsilon_{\mathrm{tr}}$ is the weight of the eigenstates
of $\rho_L$ not selected for the new block basis.
In order to perform a reliable DMRG simulation, the parameter $m$ should be
chosen such that $\epsilon_{\mathrm{tr}}$ remains small, as one further
increases the system size.
For critical 1D systems $\epsilon_{\mathrm{tr}}$
decays as a function of $m$ with a power law, while for 1D systems away
from criticality it decays exponentially, thus reflecting the entanglement
properties of the system in the two regimes: a critical system is more
entangled, therefore more states have to be taken into account.

\subsection{Finite-system DMRG}\label{ssec:finiteDMRG}

The output of the infinite-system algorithm described before is the 
(approximate) ground state of an ``infinite'' 1D chain. In other 
words, one increases the length of the chain by iterating DMRG steps, 
until a satisfactory convergence is reached. However, for many 
problems, infinite-system DMRG does not yield accurate results up to 
the wanted precision. For example, the strong physical effects of 
impurities or randomness in the Hamiltonian cannot be properly 
accounted for by infinite-system DMRG, as the total Hamiltonian is not 
yet known at intermediate steps. Moreover, in systems with strong 
magnetic fields, or close to a first order transition, one may be 
trapped in a metastable state favoured for small sizes (e.g., by edge 
effects).

Finite-system DMRG manages to eliminate such effects to a very large 
degree, and to reduce the error almost to the truncation 
error.$^{4}$ The idea of the finite-system DMRG algorithm is 
to stop the infinite-system algorithm at some preselected super-block 
length $L_{\mathrm{max}}$, which is subsequently kept fixed. In the 
following DMRG steps one applies the steps of infinite-system DMRG, 
but only one block is increased in size while the other is shrunk, 
thus keeping the super-block size constant. Reduced basis 
transformations are carried out only for the growing block.

When the infinite-system algorithm reaches the desired system size, 
the system is formed by two blocks $\mathcal B(L_{\mathrm{max}} / 2-1 , m)$
and two free sites, as shown in the first row of Fig.~\ref{fig:blocks}b.
The convergence is then enhanced by the so 
called ``sweep procedure''. This procedure is illustrated in the 
sequent rows of Fig.~\ref{fig:blocks}b. It consists in enlarging the 
left block with one site and reducing the right block correspondingly 
in order to keep the length fixed. In other words, after one 
finite-system step the system configuration is $\mathcal 
B(L_{\mathrm{max}}/2,m) \bullet \bullet \, \mathcal 
B(L_{\mathrm{max}}/2-2,m)$ (where $\bullet$ represents the free site). 
While the left block is constructed by enlarging $\mathcal 
B(L_{\mathrm{max}}/2-1,m)$ with the usual procedure, the right block 
is taken from memory, as it has been built in a previous step of the 
infinite procedure and saved. Indeed, during the initial 
infinite-system algorithm one should save the matrices $\hat{O}_{i\to 
i+1}$, the block Hamiltonians $\hat{H}_B(i)$ and the interaction 
operators $\hat{S}_i(q)$ for $i=1,L_{\mathrm{max}}/2-1$. The 
finite-system procedure goes on increasing the size of the left block 
until the length $L_{\mathrm{max}}-4$ is reached. At this stage a 
right block $\mathcal B(1,D)$ with one site is constructed from 
scratch and the left block $\mathcal B(L_{\mathrm{max}}-3, m)$ is 
obtained through the renormalization procedure. Then, the role of the 
left and right block are switched and the free sites start to sweep 
from right to left. Notice that at each step the renormalized block 
$\mathcal B(i,m_i)$ has to be stored in memory. During these sweeps 
the length of the chain does not change, thus at each step the
wavefunction of the previous one can be used as a good guess for the
diagonalization procedure (see Subsec.~\ref{subsec:guess} for details).
At each sweep the approximation of the ground state improves.
Usually two or three sweeps are sufficient to reach convergence
in the energy output.

Up to now we concentrated on a single quantum state, namely the ground
state. It is also possible to find an approximation
to a few number of states (typically less than $5$): for example,
the ground state and some low-excited state.$^{4}$
These states are called \emph{target states}. 
At each DMRG step, after the diagonalization,
for each target state $\ket{\psi_k}$ one has to calculate the corresponding
reduced density matrix $\rho_k$, by tracing the right enlarged block.
Then a convex sum of these matrices with
equal weights$^{7}$ is performed:
\begin{equation}
\rho=\frac{1}{n_k}\sum_{k=1}^{n_k} \rho_k \, .
\end{equation}
Finally $\rho$ has to be diagonalized in order to find the eigenbasis
and the transformation matrices $\hat{O}$. In this way the DMRG 
algorithm is capable of efficiently representing not only the Hilbert 
space ``around'' the ground state, but also the surroundings of the 
other target states. It is worth noting that targeting many states 
reduces the efficiency of the algorithm because a larger $m$ has to be 
used for obtaining the same accuracy. An alternative way could be to 
run as many iterations of DMRG with a single target state as many 
states are required.

\subsection{Boundary conditions}\label{ssec:boundary}

The DMRG algorithm, as it has been depicted above, describes a system 
with open boundary conditions. However, from a physical point of view, 
periodic boundary conditions are normally highly preferable to the 
open ones, as surface effects are eliminated and finite-size 
extrapolation gives better results for smaller system sizes. In the 
presented form, the DMRG algorithm gives results much less precise in 
the case of periodic boundary conditions than for open boundary 
conditions.$^{7,33,34}$
Nonetheless, periodic boundary conditions can be implemented by using the 
super-block configuration $\mathcal B \, \bullet \, \mathcal B \, 
\bullet$. This configuration is preferred over $\mathcal B \, \bullet 
\,\bullet \, \mathcal B$ because the two blocks are not contiguous,
thus enhancing, for typical situations, the sparseness of the matrices
one has to diagonalize and therefore maintaining the same computational
speed of the algorithm for open boundary conditions.$^{5}$
Simulations with the standard infinite-system super-block configuration
have also been performed in order to include twisted boundary conditions,
thus allowing the possibility to study spin stiffness or
phase sensitivity.$^{35}$

\section{Measure of observables} \label{sec:measDMRG}

Besides the energy, DMRG is also capable to extract other 
characteristic features of the target states, namely to measure the 
expectation values of a generic quantum observable $\hat{M}$. 
Properties of the $L_{\mathrm{max}}$-site system can be obtained from 
the wave functions $\ket \psi $ of the super-block at any point of the 
algorithm, although the symmetric configuration (with free sites at 
the center of the chain) usually gives the most accurate results. The 
procedure is to use the wave function $\ket\psi$ resulting from the 
diagonalization of the super-block (see the scheme in 
Sec.\ref{ssec:infiniteDMRG}, step 4), in order to evaluate expectation 
values.

We first concentrate on local observables $\hat{M} (i)$,
living on one single site $i$.
If one is performing the finite-system DMRG algorithm, it is possible to
measure the expectation value of $\hat{M} (i)$ at the particular step
inside a sweep in which $i$ is one of the two free sites.
The measure is then a simple average:
\begin{equation}
\bra\psi\hat M(i)\ket\psi = \psi^*_{a\alpha \beta b} \,
M(i)_{\alpha \alpha'} \, \psi_{a\alpha' \beta b}
\end{equation}
where $i$ is the first free site. In the special cases in which the 
observables refer to the extreme sites ($i=1$ or 
$i=L_{\mathrm{max}}$), the measurement is performed when the shortest 
block is $\mathcal B(1,D)$, following the same procedure.

It is also possible to measure an observable expectation value while 
performing the infinite-system algorithm. In this case there are two 
possibilities: either $i$ is one of the two central free sites or not. 
In the former case the measurement is performed as before, while in 
the latter one should express $\hat{M}$ in the truncated DMRG basis. 
At each DMRG iteration the operator $\hat{M} (i)$ must be updated in 
the new basis using the $\hat{O}$ matrix, as in 
Eq.~\eqref{eq:updateS}: $\hat{M} (i) \to \hat{O}^\dagger \, \hat{M} 
(i) \, \hat{O}$. The measurement is then computed as:
\begin{equation}
\langle \hat M(i)\rangle = \psi^*_{a\alpha \beta b} \, M_{a a'} \,
\psi_{a'\alpha \beta b}
\end{equation}
if site $i$ belongs to the left block and analogously if $i$ belongs 
to the right block. 

For non local observables, like a correlation function
$\hat{P} (i) \, \hat{Q} (j)$, the evaluation of expectation values
depends on whether $i$ and $j$ are on the same block or not.
The most convenient way in order to perform such type of measurements
is to use the finite-system algorithm.
Let us first consider the case of nearest neighbor observables
$\hat P(i)$ and $\hat Q(i+1)$. We can measure the expectation value
$\langle \hat P(i) \, \hat Q(i+1)\rangle$ when $i$ and $i+1$
are the two free sites.
In this case the dimensions of the matrices $\hat{P}$ and $\hat{Q}$ are
simply $(D \times D)$ and we do not have to store these operators
in block representation.
The explicit calculation of this observable is then simply:
\begin{equation}
\langle \hat P(i) \, \hat Q(i+1) \rangle = \psi^*_{a\alpha \beta b} \,
P_{\alpha \alpha'} \, Q_{\beta \beta'} \, \psi_{a\alpha' \beta' b} \, .
\end{equation}
In general, measures like $\langle \hat P(i) \, \hat Q(j)\rangle$ 
(where $i$ and $j$ are not nearest neighbor sites) can also be 
evaluated. This task can be accomplished by firstly storing the block 
representation of $\hat P(i)$ and $\hat Q(j)$, and then by performing 
the measure when $i$ belongs to a block and $j$ is a free site or 
vice-versa. Analogously, it is possible to evaluate measures in the 
case when $i$ belongs to the left block, while $j$ to the right one. 
What should be avoided is the measure of $\langle \hat P(i) \, \hat 
Q(j)\rangle$ when $i$ and $j$ belong to the \emph{same} block.
Indeed, in this case the block representation of $\hat P(i) \, \hat Q(j)$
evaluated through those of $\hat P(i)$ and $\hat Q(j)$ separately
is not correct, due to the truncation.
Instead, such type of operators have to be built up as a compound
object: in order to measure them, one has also to keep track
of the block representation of the product $\hat P(i) \, \hat Q(j)$
throughout all the calculation, consequently slowing down
the algorithm.$^{5,7}$

The standard DMRG algorithm works better with open boundary conditions
(see SubSec.~\ref{ssec:boundary}); this necessarily introduces
boundary effects in the measure of observables.
For example, in the case of spin $S=1/2$ chains, open boundaries cause
a strong alternation in the local bond strength
$\langle \hat{{\bf S}} (j) \cdot \hat{{\bf S}} (j+1) \rangle$ at the borders,
which slowly decays when shifting to the center.$^{5}$
In order to obtain a good description of the bulk system by using
open boundary conditions, one generally has to simulate a larger system
and then discard measurements on the outer sites; the number of
outer sites over which measurement outcomes strongly fluctuate
depends on the simulated physical system.

Finally, we stress that usually the convergence of measurements is 
slower than that of energy, since more finite-system DMRG sweeps are 
required in order to have reliable measurement outcomes (typically 
between five and ten). As an example, we quote the case of the
one-dimensional spin 1 Bose Hubbard model,$^{36}$ in which energies
typically converge after 2 or 3 sweeps, while the measure of the
dimerization order parameter requires at least five sweeps to converge
(the convergence gets even slower when the system approaches criticality).

\section{Time dependent DMRG} \label{sec:timeDMRG}

In this section we describe an extension of the static DMRG,
which incorporates real time evolution into the algorithm.
Various different time-dependent simulation methods
have been recently proposed,$^{10,12,13,37}$
but here we restrict our attention to the algorithm introduced by
White and Feiguin.$^{13}$

The aim of the time-dependent DMRG algorithm (t-DMRG) is to simulate
the evolution of the ground state of a nearest-neighbor one 
dimensional system described by a Hamiltonian $\hat{H}$, following the 
dynamics of a different Hamiltonian $\hat{H}_1$. In few words, the 
algorithm starts with a finite-system DMRG, in order to find an 
accurate approximation of the ground state $\ket{\psi_G}$ of $\hat{H}$. 
Then the time evolution of $\ket {\psi_G}$ is implemented, by using a 
Suzuki-Trotter decomposition$^{38,39}$ for the time 
evolution operator $\hat{U} = e^{- i \hat{H}_1 t}$.

The DMRG algorithm gives an approximation to the Hilbert subspace that
better describes the state of the system. However, during the evolution
the wave function changes and explores different parts of the Hilbert
space. Thus, the truncated basis chosen to represent the initial state 
will be eventually no more accurate. This problem is solved by 
updating the truncated bases during the evolution. The first effort, 
due to Cazalilla and Marston, consists in enlarging the effective 
Hilbert space, by increasing $m$, during the 
evolution.$^{37}$ However, this method is not very 
efficient because if the state of the system travels sufficiently far 
from the initial subspace, its representation becomes not accurate, or 
$m$ grows too much to be handled. Another solution has been proposed 
in Ref.~${13}$: the block basis should be updated at each 
temporal step, by adapting it to the instantaneous state. This can be 
done by repeating the DMRG renormalization procedure using the 
instantaneous state as the target state for the reduced density matrix.	

In order to approximately evaluate the evolution operator
$\hat{U} = e^{- i \hat{H}_1 t}$ we use a
Suzuki-Trotter decomposition.$^{38,39}$
The first order expansion in time is given by the formula:
\begin{equation} \label{1sttrotter}
e^{-i \hat{H}_1 t} \approx \left(\prod_{L=1}^{L_{\mathrm{max}}-1}
e^{-i \hat{H}_1(L,L+1) dt}\right )^n \, ,
\end{equation}
where $n=t/dt$ gives the discretization of time $t$ in small
intervals $dt$, and $\hat{H}_{L,L+1}$ is the interaction Hamiltonian
(plus the local terms) between site $L$ and $L+1$.
Further decompositions at higher orders can be obtained by
observing that the Hamiltonian can be divided in two addends:
the first, $\hat{F} = \sum_{L\, \textrm{even}} \hat{H}_1(L,L+1)$,
containing only even bonds, and the second,
$\hat{G} = \sum_{L \,\textrm{odd}} \hat{H}_1(L,L+1)$,
containing only odd bonds.
Since the terms in $\hat{F}$ and $\hat{G}$ commute,
an even-odd expansion can be performed: 
\begin{equation} \label{2ndtrotter}
e^{-i \hat{H}_1 t} \approx \left(e^{-i \hat{F} \frac{dt}{2}} \,
e^{-i \hat{G} dt} \, e^{-i \hat{F} \frac{dt}{2}} \right )^n .
\end{equation}
This coincides with a second order Trotter expansion,
in which the error is proportional to $dt^3$.
Of course, one can enhance the precision of the algorithm by using a 
fourth order expansion with error $dt^5$:$^{40}$
\begin{equation} \label{4thtrotter}
e^{-i \hat{H} t} = \prod_{i=1}^{5} \left( e^{-i p_i \, \hat{F} \,
\frac{dt}{2}} \, e^{-i p_i \, \hat{G} \, dt} \,
e^{-i \, p_i \, \hat{F} \, \frac{dt}{2}} \right)^{n}
+ O(dt^{5}) \, ,
\end{equation}
where all $p_i = 1/(4-4^{1/3})$, except $p_3 = 1 - 4 p_1 < 0$,
corresponding to evolution backward in time.

Nonetheless, the most serious error in a t-DMRG program remains the 
truncation error. A nearly perfect time evolution with a negligible 
Trotter error is completely worthless if the wave function is affected 
by a relevant truncation error. It is worth to mention that t-DMRG 
precision becomes poorer and poorer as time grows larger and larger, 
due to the accumulated truncation error at each DMRG step. This 
depends on $L_{\mathrm{max}}$, on the number of Trotter steps and, of 
course, on $m$. At a certain instant of time, called the~\emph{runaway 
time}, the t-DMRG precision decreases by several order of magnitude. 
The runaway time increases with $m$, but decreases with the number of 
Trotter steps and with $L_{\mathrm{max}}$. For a more detailed 
discussion on the t-DMRG errors and on the runaway time, see Gobert
{\it et al}.$^{41}$

The initial wave function $\ket{\psi_G}$ can be chosen from a great variety
of states. As an example, for a spin $1/2$ chain, a factorized state
can be prepared by means of space dependent magnetic fields.
In general, it is also possible to start with an initial state 
built up by transforming the ground state as
$\ket{\psi_A} = \sum_{i=1}^{L_{\mathrm{max}}} \hat{A}_i \ket{\psi_G}$,
where $\hat{A}_i$ are local operators.
The state $\ket{\psi_A}$ can be obtained by simply performing a preliminary
sweep, just after the finite-system procedure, in which
the operators $\hat{A}_i$ are subsequently applied to the transforming
wave function, when $i$ is a free site.$^{13}$

In summary, the t-DMRG algorithm is composed by the following steps:
\begin{enumerate}
\item{Run the finite-system algorithm, in order to obtain
  the ground state $\ket{\psi_G}$ of $\hat{H}$.}
\item{If applicable, perform an initial transformation 
  in order to set up the initial state $\ket{\psi_A}$.}
\item{Keep on the finite-system procedure by performing sweeps
  in which at each step the operator $e^{-i \hat{H}_1(L,L+1)dt}$
  is applied to the system state 
  ($L$ and $L+1$ are the two free sites for the current step).}
\item{Perform the renormalization, following the finite-system
  algorithm, and store the matrices $\hat{O}$ for the following steps.}
\item{At each step change the state representation to 
  the new DMRG basis using White's state prediction
  transformation$^{43}$ (see below).}
\item{Repeat points 3 to 5, until a complete $dt$ time evolution
has been computed.}
\end{enumerate}

White's state prediction transformation$^{43}$ has been firstly
developed in the framework of the finite-system DMRG to provide a good
guess for the Davidson or Lanczos diagonalization, thus enhancing the
performance of the algorithm (see Subsec.~\ref{subsec:guess} for details).
Here we briefly recall how it works, and adapt it to the
time-dependent part of the DMRG algorithm.
At any DMRG step, one has the left block $\mathcal B(L-1,m)$ and right
block $\mathcal B(L_{\mathrm{max}}-L-1,m)$ description.
To transform a quantum state $\ket\psi$ of the system in the new basis
for the next step (corresponding to the blocks $\mathcal B(L,m)$ and
$\mathcal B(L_{\mathrm{max}}-L-2,m)$) one uses the matrices $\hat{O}$: 
$\hat{O}_{L-1\to L}$ and $\hat{O}^{\dagger}_{L_{\mathrm{max}}-L-2\to 
L_{\mathrm{max}}-L-1}$. The first matrix transforms a block of length 
$L-1$ in a block of length $L$ and it has been computed in the current 
renormalization. The second one transforms a block of length 
$L_{\mathrm{max}}-L-1$ in a block of length $L_{\mathrm{max}}-L-2$; 
this matrix is recovered from memory, since it has been computed at a 
previous step. The transformed wave function then reads: 
\begin{equation} \label{whitestateprediction}
\tilde\psi_{a\alpha\beta b} = O^{*\, g(a', \alpha') \, a}_{L-1\to L} 
\: O^{g(\beta, b) \, b'}_{L_{\mathrm{max}}-L-2\to 
L_{\mathrm{max}}-L-1} \: \psi_{a' \alpha' \alpha b'} \,.
\end{equation}
Assuming this operation is already implemented, the t-DMRG algorithm
introduces only a slight modification: at step $L$ (i.e. when $L$
and $L+1$ are the two free sites), instead of the diagonalizing
the super-block with the Davidson or Lanczos, one applies
$\exp (-i \hat{H}_1 (L,L+1) dt)$ to the transformed wave function.

\begin{figure}[!ht]
  \begin{center}
    \includegraphics[scale=0.5]{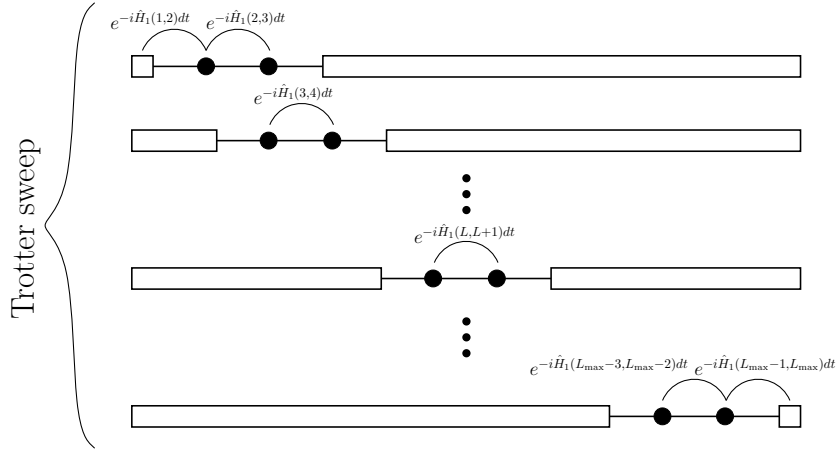}
    \caption{Schematic procedure for the t-DMRG algorithm, implemented by using
      a first order Trotter expansion for the time evolution operator.
      In this case one half sweep is needed for each time interval $dt$;
      higher order expansions require more complicated schemes,
      with an increasing number of steps.}
    \label{fig:trotter-block}
  \end{center}
\end{figure}

To compute the system time evolution using the first order Trotter expansion
of Eq.~\eqref{1sttrotter}, one should perform one half sweep for each
time interval $dt$: at the $j$-th step one has to apply
$e^{-i \hat{H}_1 (j+1,j+2) dt}$, forming the usual left-to-right sweep.
When arriving at the end of the chain, the system has been evolved of a $dt$
time; one then goes on with the next time iteration, applying the corresponding
evolution operators in a right-to-left sweep.
Attention must be paid for the border links: at the first step both
$e^{-i \hat{H}_1 (1,2) dt}$ and $e^{-i \hat{H}_1 (2,3) dt}$ have to be
applied; an analogous situation happens at the last step.
The procedure for one complete $dt$ time evolution is depicted
in Fig.~\ref{fig:trotter-block}.
Notice that, since at each step the operator $e^{-i \hat{H}_1(L,L+1)dt}$
is computed on the two current free sites $L$ and $L+1$
(or when the block is composed of just one site),
its representation is given in terms of a $D^2 \times D^2$ matrix,
and most remarkably it is exact.
More generally, if the border block dimension is such that it can be
treated exactly, it can be convenient to perform its evolution as a whole
and then switch the sweep direction.
As stated before, to increase the simulation precision, one can expand
the time evolution operator to the second order Trotter expansion,
as in Eq.~\eqref{2ndtrotter}. The implementation of this expansion
requires $3/2$ sweeps for each time interval $dt$:
in the first $e^{-i \hat{F} \frac{dt}{2}}$ is applied,
in the second $e^{-i \hat{G} dt}$, finally a third half sweep is needed
to apply $e^{-i \hat{F} \frac{dt}{2}}$ again.
In order to acquire further precision one may go to the fourth order
(see Eq.~\eqref{4thtrotter}). In this case $5 \times \frac{3}{2}$
are needed, thus the computational time is respectively five times 
or fifteen times longer than the one needed by using Eq.~\eqref{2ndtrotter}
or Eq.~\eqref{4thtrotter}.

Finally, we want to remark again that this algorithm for the time
evolution is a small modification of the finite-system procedure:
the main difference is the computation of a factor of the Trotter
expansion instead of performing the diagonalization procedure
at each step. This means that a typical t-DMRG sweep is much less
time consuming that a finite-system one. Notice also that
the measurements are performed in the same way as in the
finite-system algorithm.

\begin{figure}[!ht]
  \begin{center}
    \includegraphics[scale=0.8]{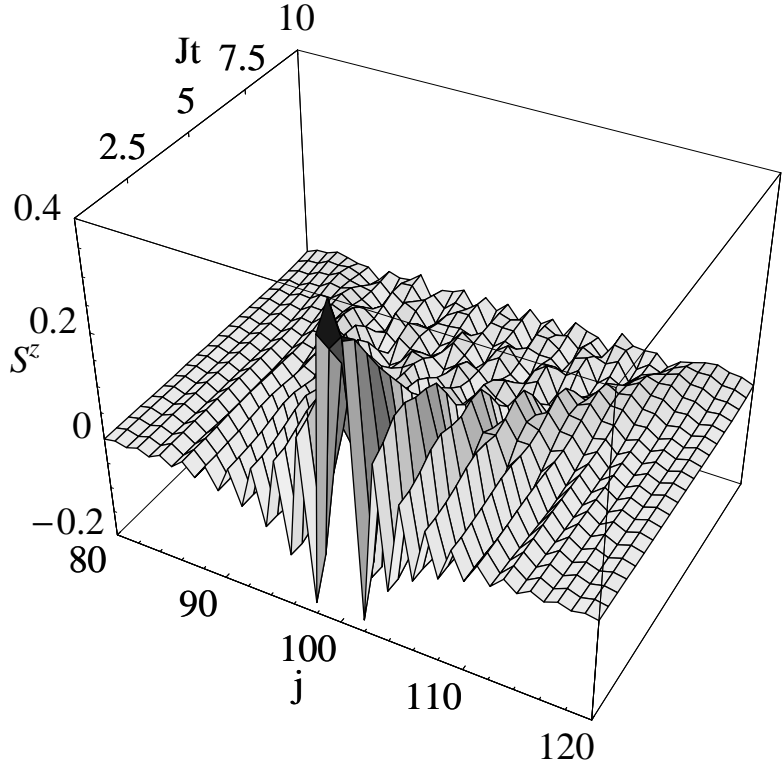}
    \caption{Temporal evolution of the local magnetization
      ${S}^z (j) $ of a 200 sites spin-1 Heisenberg chain,
      starting from the excited state obtained by applying $\hat{S}^+(100)$
      to the ground state of the chain.
      Here we used $J dt=10^{-1}$ as a Trotter slicing time, and a truncated
      Hilbert space of dimension $m=15$.}
    \label{fig:magevol}
  \end{center}
\end{figure}

To conclude this section, we provide a simple and intuitive example
which explains how the time-dependent algorithm works.
We consider the time evolution of the on-site magnetization
of an excited state for a spin-1 Heisenberg chain.$^{13}$
In order to study the dynamics of this excitation, first we run
the finite-system DMRG algorithm, thus obtaining the ground state
$\vert \psi_G \rangle$ of a $L$-sites chain.
We then perform a preliminary sweep to apply
$\hat{A} = \hat{S}^+ (j)$ on $\vert \psi_G \rangle$
for a single site $j$ located at the center of the chain,
namely we choose $\hat{A}_j = \delta_{j,L/2} \hat{S}^+ (j)$.
In this way we set up the initial state $\vert \psi_A \rangle$,
that is a localized wave packet consisting of all wave vectors.
We then perform the t-DMRG
algorithm with $\hat{H}_1$ being the Heisenberg hamiltonian, 
and instantaneously measure
the local magnetization $S^z (j)$ for each site $j$:
$\langle \psi (t) \vert \hat{S}^z (j) \vert \psi (t) \rangle$.
The initial wave packet $\vert \psi_A \rangle$ spreads out
as time progresses; different components move with different speeds,
given by the corresponding group velocity (see Fig.~\ref{fig:magevol}).

\section{Numerical examples} \label{sec:examples}

\begin{figure}[!ht]
  \begin{center}
    \includegraphics[scale=0.5]{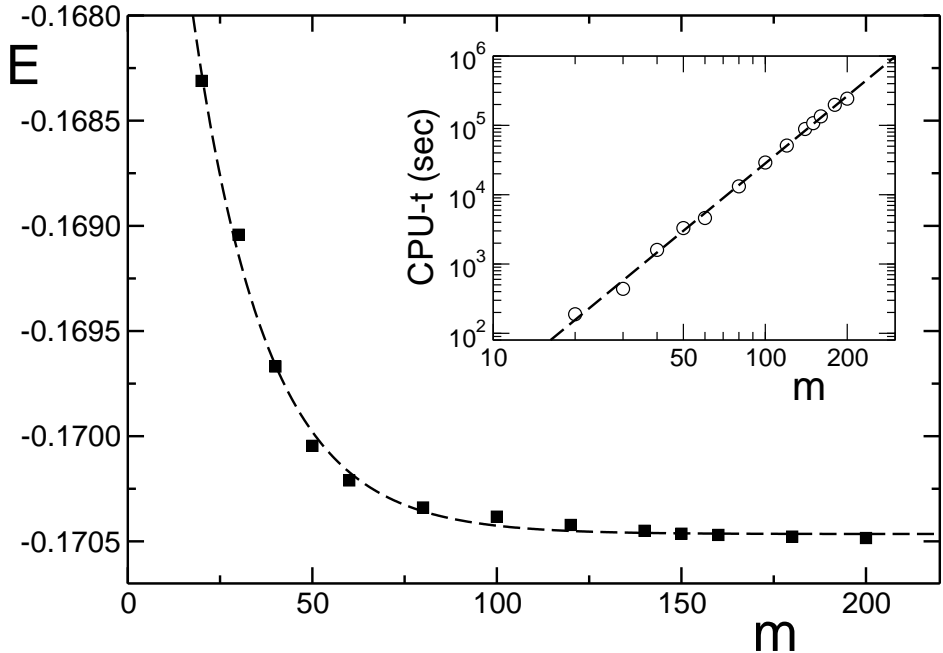}
    \caption{Ground state energy for the spinorial Bose-Hubbard model
      with a fixed number of total particles $n=L_{\mathrm{max}}$ as a
      function of $m$.
      A 1D chain of $L_{\mathrm{max}}=50$ sites has been simulated with a
      $3$-sweep finite-system algorithm. The dashed line is an exponential fit:
      $E(m)= E_0 (1 + C_0 e^{-\alpha_0 m})$ with
      $E_0 \simeq-0.17046$, $C_0 \simeq -0.035$, $\alpha_0 \simeq 0.05$.
      Inset: CPU-time dependence with $m$; dashed line shows a power
      law fit $t \sim m^\alpha$ with $\alpha \simeq 3.2$.
      \newline
      Numerical simulation presented here and in the following figures
      have been performed on a $1.6$GHz PowerPC 970 processor with
      $2.4$GB RAM memory.$^{42}$} 
    \label{fig:conv-static}
  \end{center}
\end{figure}

In this section we report some numerical examples on the convergence 
of the DMRG outputs with respect to the user fixed parameters $m$, and 
$(t, \, dt)$. Let us first focus on the static DMRG algorithm. The 
main source of error is due to the step-by-step truncation of the 
Hilbert space dimension of the system block from $m \times D$ to $m$. 
The parameter $m$ must be set up very carefully, since it 
represents the maximum number of states used to describe the system 
block. It is clear that, by increasing $m$, the output becomes closer 
and closer to the exact solution, which is eventually reached in the 
limit of $m \sim D^L$ (in that case the algorithm would no longer 
perform truncation, and the only source of error would be due to 
inevitable numerical roundoffs). As an example of the output 
convergence with $m$, in Fig.~\ref{fig:conv-static} we plotted the 
behavior of the ground state energy in the one-dimensional spin 1 
Bose-Hubbard model as a function of $m$ (see Ref.~${36}$ for a detailed 
description of the physical system). The convergence is exponential 
with $m$, as can be seen in the figure. In the inset the CPU-time 
dependence with $m$ is shown and the dashed line shows a power law fit 
of data, $m^{\alpha}$ with $\alpha \simeq 3.2$.

\begin{figure}[!ht]
  \begin{center}
    \includegraphics[scale=0.5]{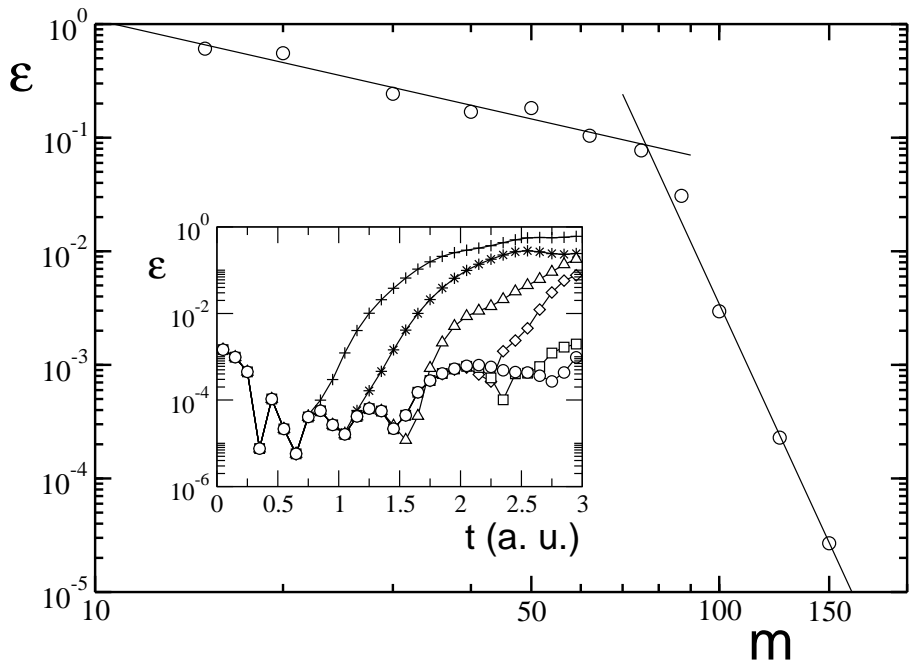}
    \caption{{Deviation $\varepsilon = |S_{\rm{exact}} - S_m|/S_m$
   at time $t=3$ as a function of $m$.
   Inset: $\varepsilon$ as a function of time for various values
   of $m$: 15 (crosses), 30 (stars), 50 (triangles),
   75 (diamonds), 100 (squares), 150 (circles).}}
    \label{fig:conv-time2}
  \end{center}
\end{figure}

We now present an example of convergence of the t-DMRG with $m$ and 
$(t, \, dt)$. We consider the dynamical evolution of the block 
entanglement entropy in a linear $XXZ$ chain (see Ref.~${20}$ for 
more details). The state of the system at time $t=0$ is the 
anti-ferromagnetic state. The initial state evolves with the 
Hamiltonian of the $XX$ model from an initial product state to an 
entangled one. This entanglement can be measured by the von Neumann 
entropy of the reduced density matrix of the block $\rho(t)$:
\begin{equation} S(t)=-\textrm{Tr} \rho(t)\log_2 \rho(t)
\end{equation}
In the example we calculate $S(t)$ for a block of size $6$ in a chain
of length $50$. The time evolution has been calculated form $t=0$ to $t=3$
with a fixed Trotter time step $dt = 5\cdot 10^{-2}$ that ensures that
the Trotter error is negligible with respect to the truncation error.  
\begin{figure}[!ht]
  \begin{center}
    \includegraphics[scale=0.5]{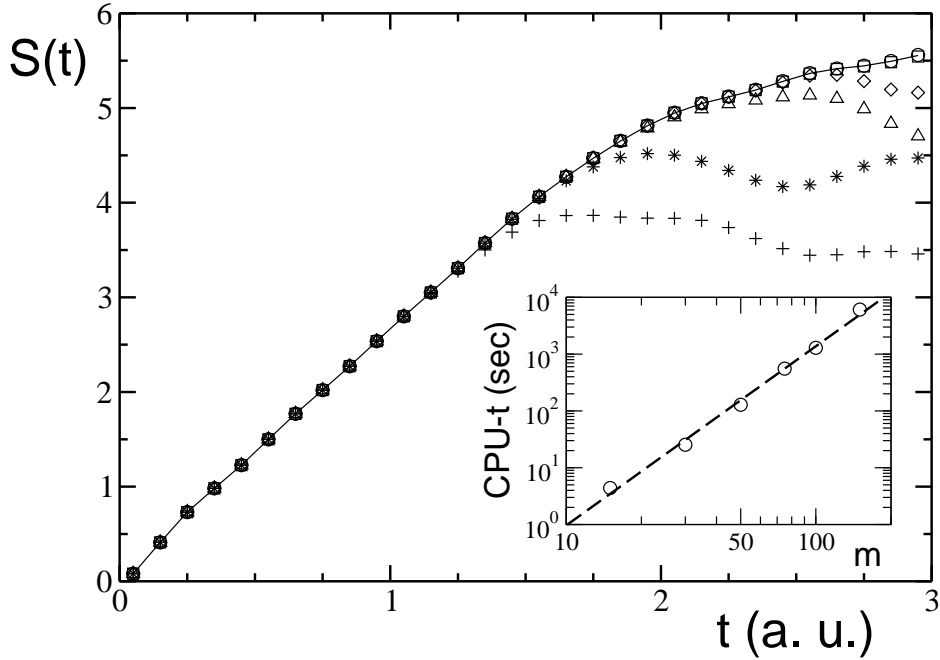}
    \caption{{Time evolution (in arbitrary units) of the von Neumann
      entropy $S(t)$ for $m=$ 15 (crosses), 30 (stars), 50 (triangles),
      75 (diamonds), 100 (squares), 150 (circles).
      The t-DMRG data are compared with the exact result (solid line).
      Inset: CPU-time dependence with $m$; dashed line shows
      a power law fit $t \sim m^\alpha$ and $\alpha \simeq 3.14$.}}
    \label{fig:conv-time1}
  \end{center}
\end{figure}
Since the $XX$ model can be solved analytically, we are able to 
compare the exact results with the t-DMRG data. In 
Fig.~\ref{fig:conv-time1} we show $S(t)$ as a function of time for 
various values of $m$, from $15$ to $150$, and compared with the exact 
data. In the inset we present the CPU-time as a function of $m$, which 
behaves as a power-law $m^\alpha$ with $\alpha \simeq 3.14$, confirming 
the estimate given in the inset of Fig.~\ref{fig:conv-static}.
The deviation $\varepsilon = |S_{\rm{exact}} - S_m|/S_m$ as a function of $m$
and of time is shown in Fig.~\ref{fig:conv-time2}.
The typical fast convergence of the DMRG result with $m$ is recovered
only when $m$ is greater than a critical value $m_c$ (two distinct
regimes are clear in Fig.~\ref{fig:conv-time2}).
This is due to the amount of entanglement present in the system:
an estimate of the number of states needed for an accurate description
is given by $m_c \propto 2^{S(t)}$.
Thus, it is always convenient to keep track of entropy to have an initial guess
for the number of states needed to describe the system.$^{41}$
On the other hand, if $m$ is increased too much, the Trotter error
will dominate and smaller $dt$ is needed to improve accuracy.

\section{Technical issues} \label{sec:techDMRG}

In this section we explain some technicalities regarding
the implementation of DMRG and t-DMRG code. They are not essential in
order to understand the algorithm, but they can be useful to anyone
who wants to write a code from scratch, or to modify the existing ones.
Some of these parts can be differently implemented, in part or
completely skipped, depending on the computational complexity
of the physical system under investigation.

\subsection{Hamiltonian diagonalization}

The ground state of the Hamiltonian is usually found by 
diagonalizing a matrix of dimensions $(m \, D)^2 \times (m \, D)^2$. 
Typically the DMRG algorithm is used when one is only interested in 
the ground state properties (at most in few low-energy eigenstates).
The diagonalization time can thus be greatly optimized by 
using Lanczos or Davidson methods: these are capable to give a small 
number ($\lesssim 10$) of eigenstates close to a previously chosen 
target energy in much less time than exact diagonalization routines. 
Moreover they are optimized for large sparse matrices, (that is the 
case of typical super-block Hamiltonians) and they do not require as 
input the full matrix. What is needed is just the effect of it on a 
generic state $\ket{\psi}$, which lives in a $(m D)^2$ dimensional Hilbert 
space. The Hamiltonian in Eq.~\eqref{htot} can be written as:
\begin{equation}
\hat{H} = \sum_p \hat{A}(p) \otimes \hat{B}(p) \, ,
\end{equation}
where $\hat{A} (p)$ and $\hat{B} (p)$ act respectively on the left
and on the right enlarged block.
Thus, only this matrix multiplication has to be implemented:
\begin{equation}
\psi^{\rm{out}}_{a\alpha \beta b} = \sum_{p} \hat A(p)^{\, g(a,\alpha) 
\, g(a',\alpha')} \: \hat B(p)^{\, g(b,\beta) \, g(b',\beta')} \: 
\psi^{\rm{in}}_{a'\alpha'\beta' b'}
\end{equation}
In this way it is possible to save a great amount of memory
and number of operations, since the dimensions of $\hat{A} (p)$ and
$\hat{B} (p)$ are $(m \, D) \times (m \, D)$,
and not $(m \, D)^2 \times (m \, D)^2$.
As an example, a reasonable $m$ value for simulating the evolution
of a $L_{\mathrm{max}} = 50$ spin $1/2$ chain ($D=2$) is $m \sim 50$.
This means that, in order to store all the $\sim 10^8$ complex
numbers of $\hat{H}_{supB}$ in double precision,
$\sim 1.6$ Gbytes of RAM is needed.
Instead, each of the two matrices $\hat{A}$ and $\hat{B}$ requires
less than $200$ kbytes of RAM.

\subsection{Guess for the wave function} \label{subsec:guess}

Even by using the tools described in the previous paragraph, the most
time consuming part of a DMRG step remains the diagonalization procedure.
The step-to-step wave function transformation required for the
t-DMRG algorithm, which has been described in the previous section,
can also be used in the finite-system DMRG, in order to speed up
the super-block diagonalization.$^{43}$
Indeed the Davidson or Lanczos diagonalization methods are iterative algorithms
which start from a generic wave function, and then recursively modify it,
until the eigenstate closest to the target eigenvalue is reached
(up to some tolerance value, fixed from the user).
If a very good initial guess is available for the diagonalization procedure,
the number of steps required to converge to the solution can be drastically
reduced and the time needed for the diagonalization can be reduced 
up to an order of magnitude.

In the finite-system algorithm the system is changing much less than
in the infinite algorithm, and an excellent initial guess is found to be
the final wave function from the previous DMRG step,
after it has been written in the new basis for the current step.
White's prediction is used in order to change the basis of the previous
ground state with the correct operators $\hat{O}$,
as in Eq.~\eqref{whitestateprediction}.
It is also possible to speed up the diagonalization in the infinite-system
algorithm, but here the search for a state prediction is
slightly more complicated (see, e.g., Refs.~${44,45}$).

\subsection{Symmetries}

\begin{figure}[htbp]
  \begin{center}
    \includegraphics[scale=1]{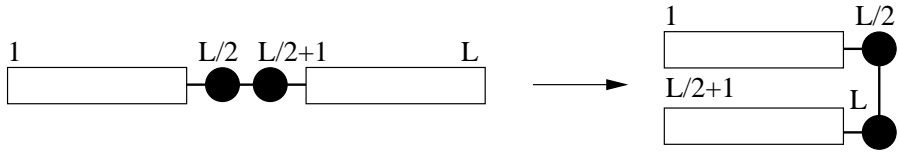}
    \caption{Alternative labelling of sites, to be used in the
      environment reflection procedure (in case of globally
      reflection-symmetric systems).}
    \label{fig:reflection}
  \end{center}
\end{figure}

If the system has a global reflection symmetry, it is possible to take 
the environment block equal to the system block, in the 
infinite-system procedure. Namely, the right enlarged block is simply 
the reflection of the left one. To avoid the complication of the 
reflection we can consider an alternative labelling of the sites, as 
shown in Fig.~\ref{fig:reflection}. In this case left and right 
enlarged blocks are represented by exactly the same matrix.

If other symmetries hold, for example conservation of angular momentum
or particle number, it is possible to take advantage of them, such to
considerably reduce the CPU-time for diagonalization. The idea is to 
rewrite the total Hamiltonian in a block diagonal form, and then 
separately diagonalize each of them. If one is interested in the 
ground state, he simply has to compare the ground state energies 
inside each block, in order to find the eigenstate corresponding to 
the lowest energy level. One may also be interested only in the ground 
state with given quantum numbers (for example in the Bose-Hubbard 
model one can fix the number of particles);
in this case one has to diagonalize the block Hamiltonian
corresponding to the wanted symmetry values.
In order to perform this task, it is sufficient to divide the operators
for the left and right block into different symmetry sectors.
Then the multiplication will take into account only the sectors'
combination which preserves the total quantum number.
When finding the reduced density matrix $\rho$, its eigenstates
have also to be symmetry-labelled.
Attention must be paid when truncating to the reduced basis:
it is of crucial importance to retain whole blocks of eigenstates
with the same weight, inside a region with given quantum numbers.
This helps in avoiding unwanted artificial symmetry breaking,
apart from numerical roundoff errors.

\subsection{Sparse Matrices}

Operators typically involved in DMRG-like algorithms (such as block 
Hamiltonians, updating matrices, observables) are usually represented 
by sparse matrices. A well written programming code takes advantage of 
this fact, thus saving large amounts of CPU-time and memory. Namely, 
there are standard subroutines which list the position (row and 
column) and the value of each non null element for a given sparse matrix.

\subsection{Storage}

Both the static and the time dependent DMRG require to store a great 
number of operators: the block Hamiltonian, the updating matrices, and 
if necessary the observables for each possible block length. One 
useful way to handle all these operators is to group each of them in a 
register, in which one index represents the length of the block. 
Operatively, we store all these operators in the fast-access RAM 
memory. However, for very large problems one can require more than the 
available RAM, therefore it is necessary to store these data in the 
hard disk. The read/write operations from hard disk have to be
carefully implemented, e.g., by performing them asyncronously, since
a non optimal implementation may dramatically slow down
the program performance.

\subsection{Algorithm Schemes}

\begin{figure}
  \begin{center}
    \includegraphics[scale=0.72]{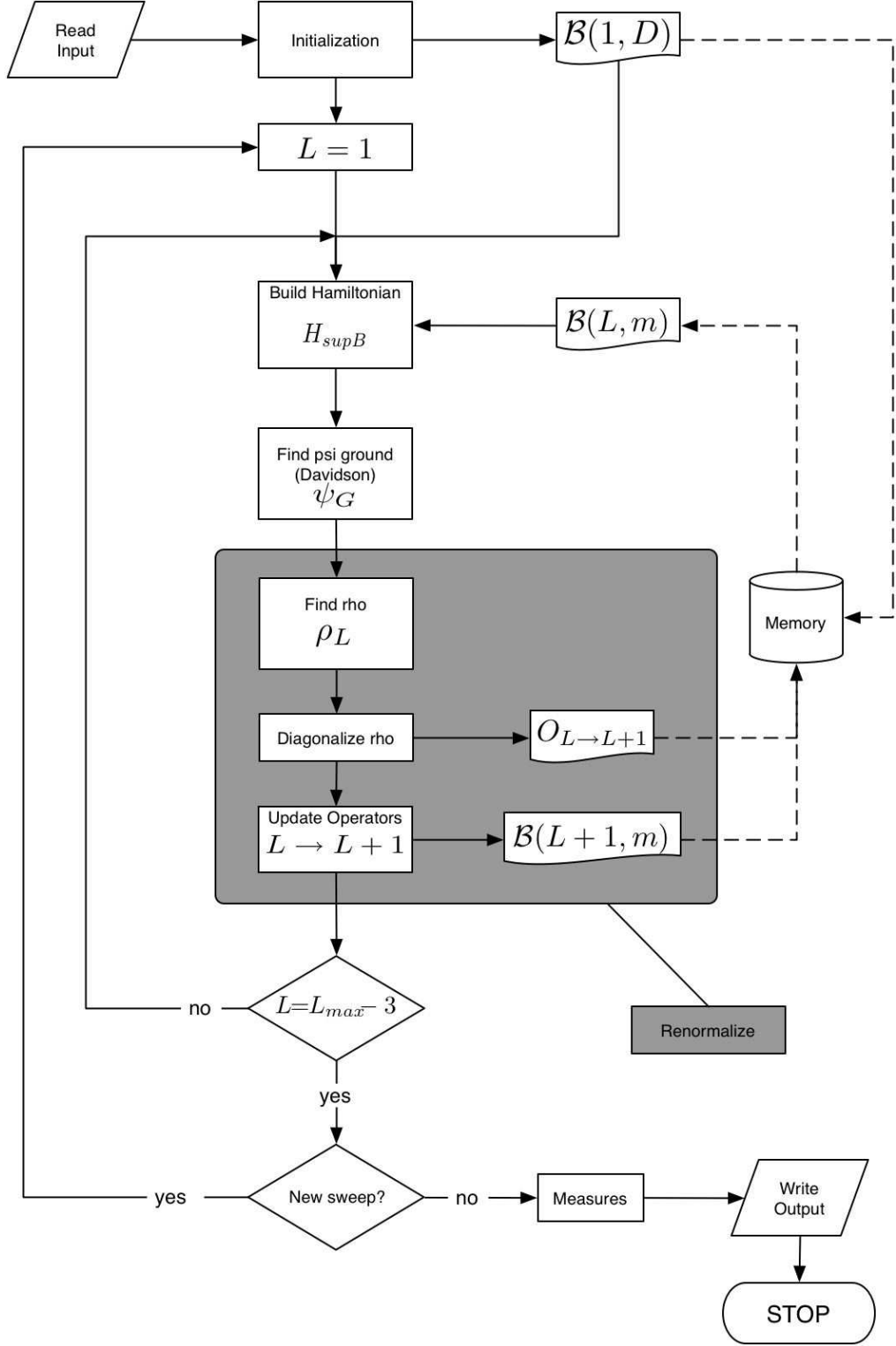}
    \caption{Basic scheme of the infinite/finite DMRG algorithm. Here we have
      supposed, for simplicity, that the system is globally reflection
      symmetric (thus the environment block is taken equal to the system
      block). The shadowed rectangle is the basic renormalization stage
      that will be used also in the t-DMRG algorithm
      (see Fig.~\ref{fig:schemat}).
      Notations are the same as those used in the text.}
    \label{fig:schema}
  \end{center}
\end{figure}

\begin{figure}
  \begin{center}
    \includegraphics[scale=0.8]{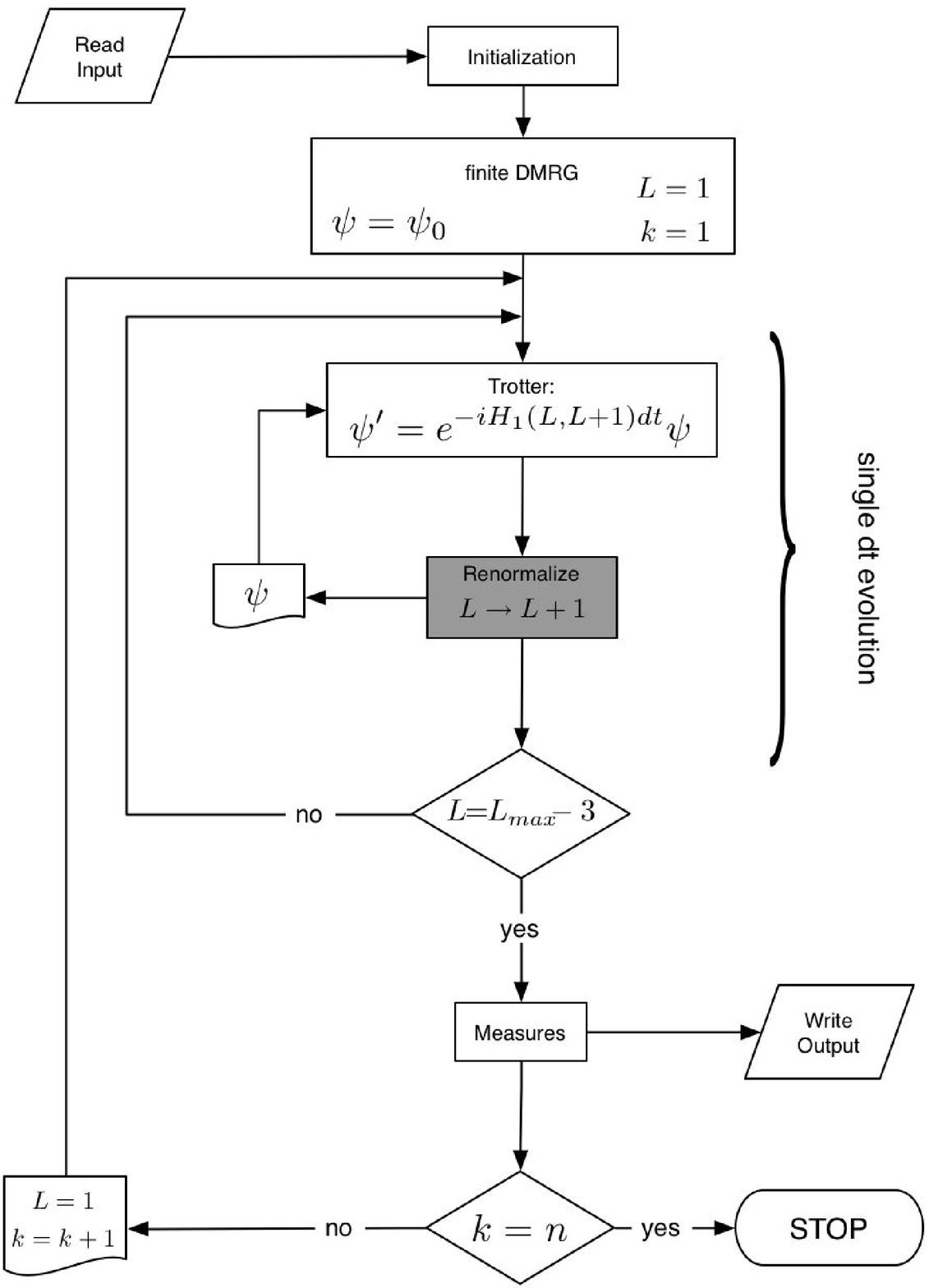}
    \caption{Basic scheme of the time-dependent DMRG algorithm.
      The index $k$ refers to the discretized time and $n=t/dt$.}
    \label{fig:schemat}
  \end{center}
\end{figure}

Figures~\ref{fig:schema}-\ref{fig:schemat} show a flow-chart
schematic representation of DMRG and t-DMRG code.

\section*{Acknowledgments}

We thank J.J. Garcia-Ripoll and C. Kollath for some technical comments 
while developing the DMRG code and for feedback. We also thank G.L.G. 
Sleijpen$^{32}$ for providing us with the Fortran routine to find 
the ground state. This work has been performed within the ``Quantum 
Information'' research program of Centro di Ricerca Matematica ``Ennio 
de Giorgi'' of Scuola Normale Superiore and it has been partially 
supported by IBM 2005 Scholars Grants -- Linux on Power and by NFS 
through a grant from the ITAMP at Harvard University and the 
Smithsonian Astrophysical Observatory. SM acknowledges support from 
the Alexander Von Humboldt Foundation. GDC acknowledges support from the
European Commission through the FP6-FET Integrated Project SCALA, CT-015714.

\section*{REFERENCES}

{\bf 1.}
  R. Feynman,
  Int. J. Theor. Phys. {\bf 21}, 467 (1982).

{\bf 2.}
  See, e.g., {\it A Guide to Monte Carlo Simulations in Statistical Physics},
  D.P. Landau and K. Binder, Cambridge University Press (2000).
  A.W. Sandvik and J. Kurkij\"arvi, Phys. Rev. B {\bf 43}, 5950 (1991).

{\bf 3.}
  H.A. van der Vorst,
  {\it Computational Methods for large Eigenvalue Problems},
  in P.G. Ciarlet and J.L. Lions (eds), 
  Handbook of Numerical Analysis, Volume VIII, 
  North-Holland (Elsevier), Amsterdam (2002).

{\bf 4.}
  S.R. White,
  Phys. Rev. Lett. {\bf 69}, 2863 (1992);

{\bf 5.}
  S.R. White,
  Phys. Rev. B {\bf 48}, 10345 (1993).

{\bf 6.}
  I. Peschel, X. Wang, M. Kaulke, and K. Hallberg, eds.,
  {\it Density-Matrix Renormalization} (Springer Verlag, Berlin, 1999).

{\bf 7.}
  U. Schollw\"ock,
  Rev. Mod. Phys. {\bf 77}, 259 (2005).

{\bf 8.}
  K.G. Wilson,
  Rev. Mod. Phys. {\bf 47}, 773 (1975).

{\bf 9.}
  S. \"Ostlund and S. Rommer,
  Phys. Rev. Lett. {\bf 75}, 3537 (1995).

{\bf 10.}
  G. Vidal,
  Phys. Rev. Lett. {\bf 91}, 147902 (2003).

{\bf 11.}
  G. Vidal,
  Phys. Rev. Lett. {\bf 93}, 040502 (2004).

{\bf 12.}
  A.J. Daley, C Kollath, U. Schollw\"ock, and G. Vidal,
  J. Stat. Mech. P04005 (2004).

{\bf 13.}
  S.R. White and A.E. Feiguin,
  Phys. Rev. Lett. {\bf 93}, 076401 (2004).

{\bf 14.}
  A.E. Feiguin and S.R. White,
  Phys. Rev. B {\bf 72}, 020404 (2005).

{\bf 15.}
  S.R. Manmana, A. Muramatsu, and R.M. Noack,
  AIP Conf. Proc. {\bf 789}, 269 (2005), arXiV:cond-mat/0502396.

{\bf 16.}
  J.J. Garcia-Ripoll,
  New J. Phys. {\bf 8}, 305 (2006).

{\bf 17.}
  U. Schollw\"ock and S.R. White,
  in G.G. Batrouni and D. Poilblanc (eds), Effective models for low-dimensional
  strongly correlated systems, p. 155, AIP, Melville, New York (2006).

{\bf 18.}
  G. Vidal, J.I. Latorre, E. Rico, and A. Kitaev,
  Phys. Rev. Lett. {\bf 90}, 227902 (2003).
  J.I. Latorre, E. Rico, and G. Vidal,
  Quant. Inf. and Comp. {\bf 4}, 48 (2004).

{\bf 19.}
  P. Calabrese and J. Cardy,
  J. Stat. Mech.: Theor. Exp. P04010 (2005).

{\bf 20.}
  G. De Chiara, S. Montangero, P. Calabrese, and R. Fazio,
  J. Stat. Mech.: Theor. Exp. P03001 (2006).

{\bf 21.}
  R. Orus and J.I. Latorre,
  Phys. Rev. A {\bf 69}, 052308 (2004).

{\bf 22.}
  S. Montangero,
  Phys. Rev. A {\bf 70}, 032311 (2004).

{\bf 23.}
  S.Bose,
  Phys. Rev. Lett. {\bf 91}, 207901 (2003).

{\bf 24.}
  D. Rossini, T. Calarco, V. Giovannetti, S. Montangero, and R. Fazio,
  Phys. Rev. A {\bf 75}, 032333 (2007).

{\bf 25.}
  A. Osterloh, L. Amico, G. Falci, and R. Fazio,
  Nature (London) {\bf 416}, 608 (2002).

{\bf 26.}
  K. Hallberg,
  Adv. Phys. {\bf 55}, 477 (2006).

{\bf 27.}
  G. Vidal,
  arXiV:cond-mat/0512165.

{\bf 28.}
  A.L. Malvezzi,
  Braz. J. of Phys. {\bf 33}, 55 (2003).

{\bf 29.}
  R.M. Noack and S.R. Manmana,
  AIP Conf. Proc. {\bf 789}, 93 (2005), arXiV:cond-mat/0510321.

{\bf 30.}
  If mirror symmetry does not hold
  the right enlarged block must be built up independently.

{\bf 31.}
  If $m$ is bigger than $D^2$ we do not truncate the basis at the first step.
  The truncation starts when $m < D^L$ and $L$ is the number of spins
  in the enlarged block.

{\bf 32.}
  To find the ground state we used the JDQZ routine implemented
  by D. Fokkema, G.L.G. Sleijpen, and H.A. van der Vorst
  \texttt{http://www.math.ruu.nl/people/sleijpen/} .

{\bf 33.}
  M.C. Chung and I. Peschel,
  Phys. Rev. B {\bf 62}, 4191 (2000).

{\bf 34.}
  F. Verstraete, D. Porras, and J.I. Cirac,
  Phys. Rev. Lett. {\bf 93}, 227205 (2004).

{\bf 35.}
  P. Schmitteckert and U. Eckern,
  Phys. Rev. B {\bf 53}, 15397 (1996).

{\bf 36.}
  M. Rizzi, D. Rossini, G. De Chiara, S. Montangero, and R. Fazio,
  Phys. Rev. Lett. {\bf 95}, 240404 (2005).

{\bf 37.}
  M.A. Cazalilla and J.B. Marston,
  Phys. Rev. Lett. {\bf 88}, 256403 (2002);
  Phys. Rev. Lett. {\bf 91}, 049702 (2003).

{\bf 38.}
  M. Suzuki,
  Prog. Theor. Phys. {\bf 56}, 1454 (1976).

{\bf 39.}
  H.F. Trotter,
  Proc. Am. Math. Soc. {\bf 10}, 545 (1959).

{\bf 40.}
  U. Schollw\"ock,
  J. Phys. Soc. Jpn. {\bf 74} (Suppl.), 246 (2005).

{\bf 41.}
  D. Gobert, C. Kollath, U. Schollw\"ock, and G. Sch\"utz,
  Phys. Rev. E {\bf 71}, 036102 (2005).

{\bf 42.}
  See for example:
  \texttt{http://www-03.ibm.com/servers/eserver/bladecenter/} .

{\bf 43.}
  S.R. White,
  Phys. Rev. Lett. {\bf 77}, 3633 (1996).

{\bf 44.}
  U. Schollw\"ock,
  Phys. Rev. B {\bf 58}, 8194 (1998).

{\bf 45.}
  L. Sun, J. Zhang, S. Qin, and Y. Lei,
  Phys. Rev. B {\bf 65}, 132412 (2002).

\end{document}